# Monolithic integration of single quantum emitters in hBN bullseye cavities

Lesley Spencer[1,2], Jake Horder[1], Sejeong Kim[3], Milos Toth[1,2] and Igor Aharonovich[1,2*]

1. School of Mathematical and Physical Sciences, University of Technology Sydney, Ultimo, New South Wales 2007, Australia
2. ARC Centre of Excellence for Transformative Meta-Optical Systems, University of Technology Sydney, Ultimo, New South Wales 2007, Australia
3. Department of Electrical and Electronic Engineering, University of Melbourne, Parkville, Victoria 3052, Australia

milos.toth@uts.edu.au; igor.aharonovich@uts.edu.au

**Abstract.**
The ability of hexagonal boron nitride to host quantum emitters in the form of deep-level color centers makes it an important material for quantum photonic applications. This work utilizes a monolithic circular Bragg grating device to enhance the collection of single photons with 436 nm wavelength emitted from quantum emitters in hexagonal boron nitride. We observe a 6-fold increase in collected intensity for a single photon emitter coupled to a device compared to an uncoupled emitter, and show exceptional spectral stability at cryogenic temperature. The devices were fabricated using a number of etching methods, beyond standard fluorine-based reactive ion etching, and the quantum emitters were created using a site-specific electron beam irradiation technique. Our work demonstrates the potential of monolithically-integrated systems for deterministically-placed quantum emitters using a variety of fabrication options.

Quantum states of light are highly sought after for emerging technologies offering accelerated computation [1], secure communication [2], and enhanced sensing [3]. Applications which make use of single photon number states in particular place stringent performance requirements on the single photon sources (SPSs) [4]. Namely, the source should output a train of single, indistinguishable photons at a rate on the order of 1 GHz. While single photon purity and photon indistinguishability characterize the distinction between quantum and classical light, the photon count rate, or SPS brightness, is a key performance metric for technologies like photonic quantum computation and quantum key distribution [5]. Solid state SPSs have been established as an attractive photonics platform for these technologies [6], particularly lattice defects in wide bandgap semiconductors such as diamond and hexagonal boron nitride (hBN) which are known to host bright, stable emitters at room temperature [7–9]. The layered nature of hBN allows for simple exfoliation and transfer methods which enable versatile sample preparation and device integration techniques [10,11]. Several approaches to defect-based quantum emitter creation in hBN have been explored, including annealing [12,13], electron beam irradiation [14–16], and ion beam irradiation [17,18]. Quantum emitters in bulk hBN are typically characterized by an in-plane dipole emission pattern and have an excited state lifetime on the order of 2 ns [19,20], which corresponds to a maximum photon emission rate of 0.5 GHz.

The photon detection rate can be increased toward the emission rate by enhancing the photon collection efficiency through radiation mode shaping, and/or by reducing the lifetime through Purcell enhancement. The circular Bragg grating (CBG) has been a popular approach to address this need. Enhanced collection efficiency from CBG structures has been reported for various SPS platforms, including self-assembled quantum dots (QDs) [21–25], colloidal QDs [26,27], bulk diamond [28] and nanodiamond [29], and CBGs have also been used to enhance excitonic bandgap emission [30,31]. In hBN, ensemble emission from the boron vacancy spin defect was enhanced using monolithic integration with a CBG device [32], and the recently discovered 436 nm quantum emitter, termed B-center, was coupled to a waveguide and the emission efficiently collected using a semicircular Bragg-style output coupler [33]. Early works on QDs and diamond etched the CBG structure into an existing host material containing a nominal density of single photon emitters, resulting in a low probability of cavity coupling. Deterministic positioning has since been achieved using pick-and-place methods that require secondary dielectric deposition to fully embed the emitter within the cavity, and can still suffer from random out-of-plane dipole orientations. In contrast, single B-centers have in-plane dipoles and they can be positioned deterministically relatively easily in pre-fabricated hBN photonic structures [34], making them uniquely suited to monolithic integration with CBG devices. Here we enhance the collection efficiency of photons generated by a single B-center using a monolithic CBG in hBN. The emitter brightness is found to be sensitive to the CBG dimensions, and for an optimal design we observe a high spectral stability, suggesting that the employed electron beam lithography (EBL) and reactive ion etching (RIE) fabrication methods impart negligible damage to the hBN lattice in the vicinity of the emitters.

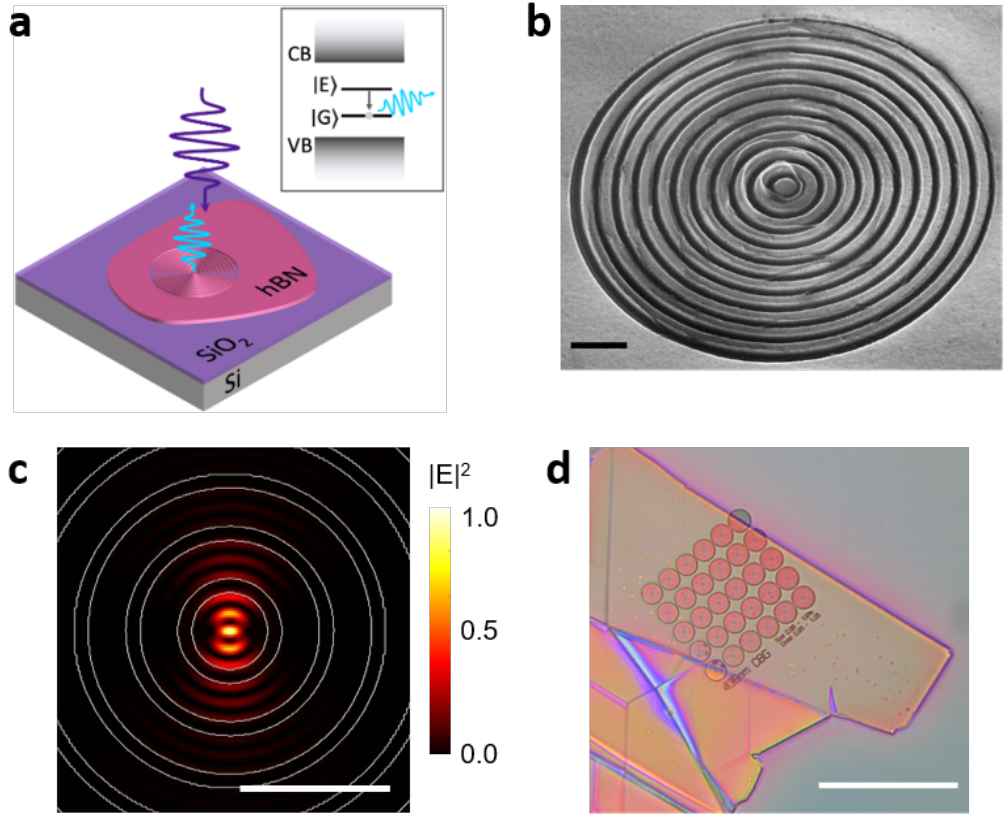

***Figure 1. Circular Bragg grating cavity.*** *a) Schematic of a B-center in an hBN CBG on a SiO$_2$/Si substrate, excited by a 405 nm laser. The inset shows the energy level structure of the emitter: CB, conduction band; VB, valence band; |G⟩, defect ground state; |E⟩, defect excited state. b) SEM image of an hBN CBG cavity. The scale bar is 1 μm. c) Simulated in-plane normalized electric field intensity from a dipole source embedded in the CBG cavity center. The scale bar is 1 μm. d) Optical microscope image of an array of CBG devices patterned in resist using EBL. The scale bar is 50 μm.*

In our previous work [32], we designed a CBG structure optimized for boron vacancy ensembles, which are characterized by a broad emission spectrum centered at 780 nm. The geometry of that successful design has been rescaled to accommodate the higher energy 436 nm photons produced by the B-center, shown schematically in Figure 1a. The SEM image in Figure 1b shows the resulting CBG, with a central disk diameter of 512 nm, ring width of 256 nm, and ring spacing of 97 nm. Single B-centers can be created within the central disk easily using electron beam irradiation, although their in-plane orientation is random. The CBG cavity mode is simulated in Figure 1c, illustrating normalized electric field intensity. The cavity is designed to induce resonance at 436 nm so that it matches the wavelength of the B-center. The cavity is resilient to variations in the in-plane orientation of the dipole due to the structural rotational symmetry of the CBG. Deterministic positioning of B-centers allowed dense arrays of CBGs to be fabricated, as shown in Figure 1d, in order to systematically investigate the influence of fabrication parameters.

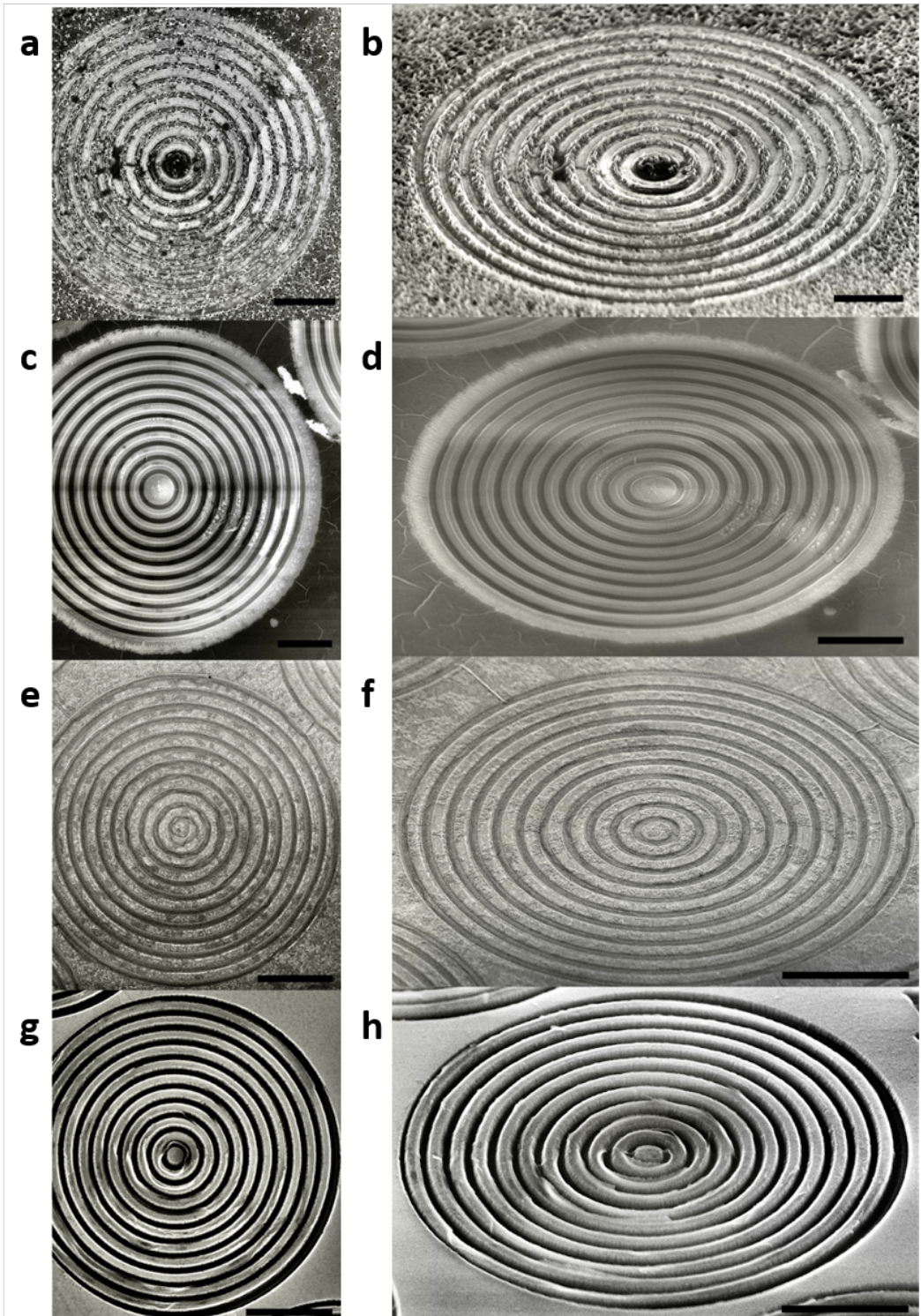

*Figure 2. SEM images of CBG structures fabricated using several etching methods.* The images are at 0° tilt (left column) and 52° tilt (right column): (a,b) argon IBE with PMMA mask; (c,d) argon and sulfur hexafluoride CAIBE with PMMA mask; (e,f) argon and chlorine RIE with CSAR mask removed; (g,h) argon and sulfur hexafluoride RIE with CSAR mask removed. Each scale bar is 2 μm.

One such parameter that we explored is the dry etching method used to transfer the EBL pattern into the hBN. Photonic structures in hBN are commonly etched by fluorine-based RIE [33,35,36]. Here, we compare this approach to initial tests of chlorine RIE, ion beam etching (IBE), and chemically-assisted ion beam etching (CAIBE). In all cases, we used the EBL resist as a hard mask for the etching process. This is done to eliminate fabrication imperfections caused by the liftoff process required for a metal mask, which we avoid at the expense of mask robustness.

Figure 2a,b show a CBG fabricated using a five minute argon IBE etch process, using a PMMA mask. The PMMA was ineffective as a mask for the physical argon etch process and appears to have been significantly sputtered whilst the exposed hBN is only slightly etched. The etched hBN has rough surfaces and slanted sidewalls from the mask being sputtered away throughout the etch. In comparison, Figure 2c,d show a CBG fabricated using a ten minute argon CAIBE etch. Similar to the IBE process, this is a physical argon etch, but it is chemically assisted by sulfur hexafluoride ($SF_6$) gas injected at the hBN surface. We note that the surface roughness of both the mask and hBN is substantially lower than that generated by the IBE etch. However, the CAIBE process caused cracking of the polymer mask and lateral etching near regions that were developed leading to the slanted sidewalls, similar to those observed in the IBE etch. The PMMA mask appears to have withheld longer under the CAIBE conditions, leading to a longer etch time and resulting in a deeper etch. Both the IBE and CAIBE methods can therefore be improved by the use of a more robust mask and a chemical etch component.

Next, in Figure 2e and Figure 2f we present SEM images of a CBG fabricated using a chlorine-based RIE etch. This sample was etched for two minutes with a CSAR mask that was chemically removed with CSAR remover before SEM imaging. This etch is shallow but appears to have the most precise features with the smoothest sidewalls. Compared to the fluorine etch, shown in Figures 2g and 2h after chemical CSAR mask removal, the chlorine-based RIE etch as well as the IBE and CAIBE etches are very slow. The fluorine recipe etches through the 300 nm flake in a third of the time that the chlorine etch was run. However, it shows a roughness of the sidewalls known as microtrenching. This is an undesired result of RIE that degrades the photonic properties of a structure, especially since a given physical defect becomes more significant as the features of the structure become smaller and approach a scale similar to the defect. Despite this, our fluorine-based etch produced the devices most appropriate for coupling to emitters and characterisation due to the etch depth. Whilst the emitter coupling and characterisation for the rest of this work will be conducted with the devices fabricated with fluorine based RIE, we acknowledge the potential for development of each etching method and refinement of fluorine-based RIE.

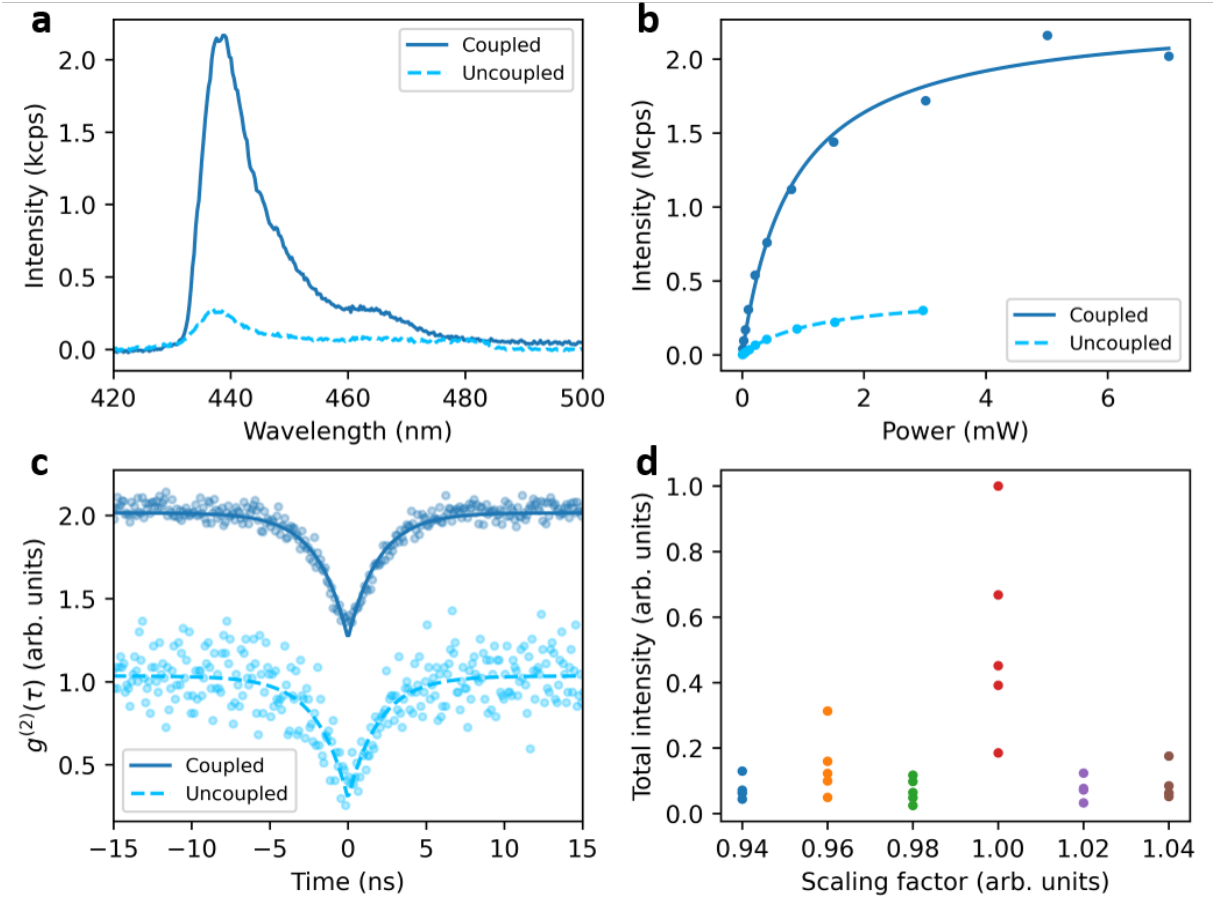

*Figure 3. Coupling single emitters to CBG devices.* a) PL spectra from a B-center emitter within a CBG and within bulk hBN, using 500 μW of 405 nm excitation over 10 s. b) Power saturation behavior for each emitter in (a). c) Second order correlation measurement for each emitter in (a), using 500 μW of 405 nm excitation. d) Integrated spectral emission intensity from emitters placed in a CBG with varying lattice constants.

Continuing now with the fluorine RIE fabrication, the sensitivity of the design dimensions presented in Figure 1b was evaluated by comparing the intensity of B-centers placed in CBG devices fabricated with varying spatial and lithographic parameters, as well as the comparison to B-centers in bulk, unstructured hBN. This latter case is shown in Figure 3a, where the bright photoluminescence (PL) spectrum from an emitter coupled to a CBG device is clearly distinguished from the spectrum of the same type of emitter situated in bulk hBN (uncoupled). Under the same excitation conditions of 500 μW of 405 nm laser, the CBG-coupled emitter has a six times greater total intensity than the uncoupled emitter. The increased spectral intensity is attributed primarily to out-of-plane emission redirection, rather than lifetime reduction, since we expect the CBG to have negligible quality factor and hence minimal Purcell enhancement.

The coupling to the CBG is also evident in the power saturation behavior, as shown in Figure 3b. Here, the observed saturation power for the coupled and uncoupled emitters are of similar magnitude, being 1.23 mW and 0.82 mW respectively, but we find a six-fold increase in the saturation intensity from 0.4 Mcps with the uncoupled emitter to 2.3 Mcps for the coupled emitter. Hence, for the same excitation power the CBG allows for a much greater proportion of emitted photons to be collected, and also reduces the relative proportion of background fluorescence. The second order correlation measurements in Figure 3c indicate the single photon nature of the emission from the two emitters under study, further evidence that increased intensity is due to the action of the CBG structure rather than the presence of multiple single photon emitters.

To compare the performance of each CBG, the total emission intensity was calculated by summing the PL spectra data. The influence of device geometry is shown in Figure 3d, where total intensity data are grouped by variation of scaling factor, relative to the dimensions of the device in Figure 1b. The initial design is seen to be optimal, since the average intensity is rapidly reduced for small increments of the scaling factor in either direction.

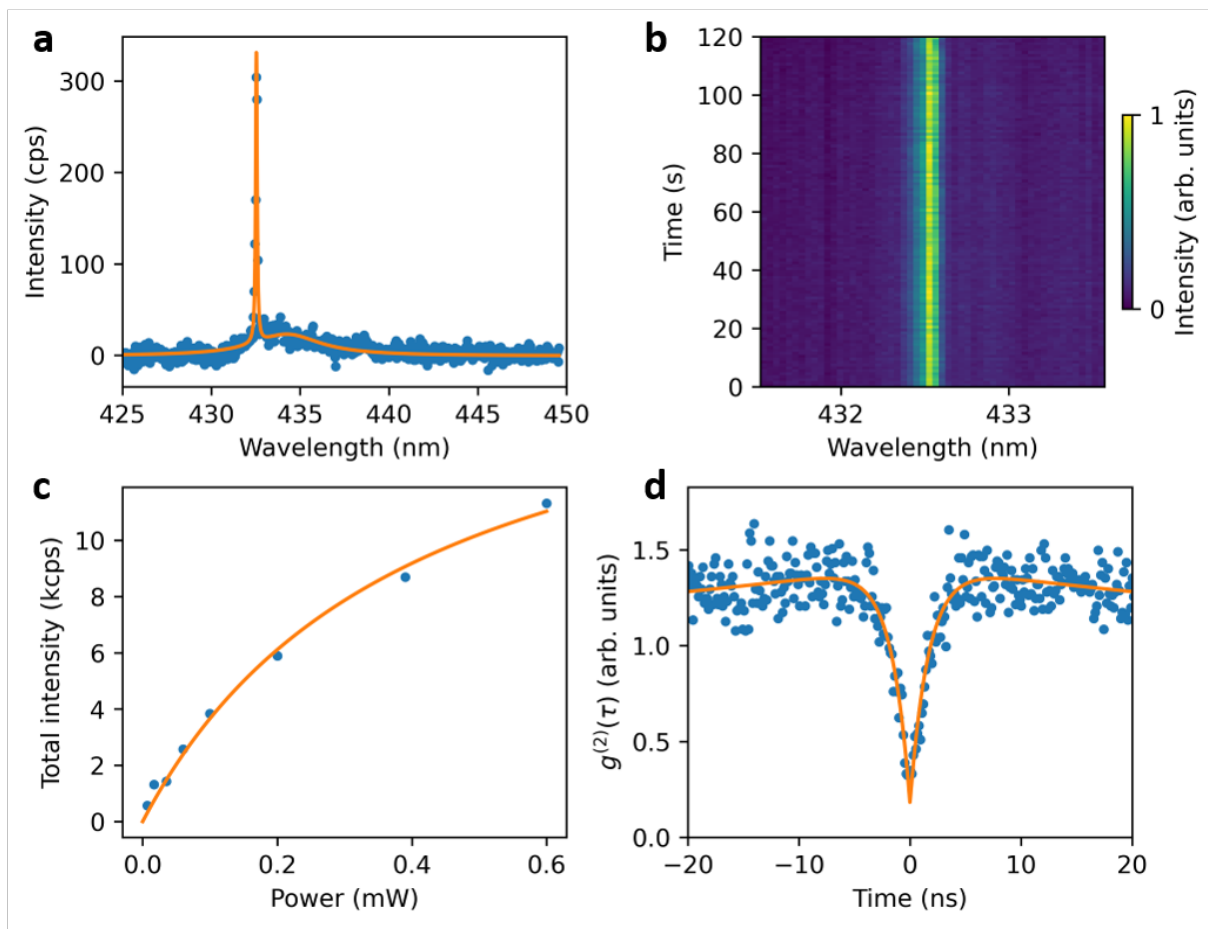

*Figure 4. Cryogenic spectroscopy and photon statistics. a) PL spectrum from an emitter monolithically embedded in a CBG device, using a 405 nm excitation laser at 800 µW power and integrating for 1 s with an 1800 l/mm grating. The fit in orange is composed of two Lorentzian functions. b) Kinetic PL spectrum over 2 minutes, showing the stability of the ZPL. c) Power saturation behavior, using the total intensity of spectra integrated for 5 s with an 1800 l/mm grating. The fit in orange yields a saturated emission intensity of $I_\infty$ = 18.47 thousand counts/s, and $P_{sat}$ = 0.40 mW. d) Second order correlation measurement, using 800 µW of 405 nm laser over 60 minutes. The fit in orange yields a minimum of $g^{(2)}(0)$ = 0.18.*

Next, we cooled the CBG array sample to 5 K and performed cryogenic spectroscopy with a well-coupled emitter. Off-resonant excitation reveals a bright, spectrometer-limited zero phonon line (ZPL) accompanied by a low energy acoustic phonon sideband, as shown in Figure 4a. Fitting both features with a Lorentzian curve yields a ZPL full width at half maximum (FWHM) of 0.1 nm and PSB FWHM of 5 nm. The proximity of this sideband introduces significant dephasing that ultimately limits the visibility in two photon interference experiments [37], although moderate Hong-Ou-Mandel visibility was recently achieved with a B-center under off-resonant excitation [38]. Photon indistinguishability would be improved with narrow filtering of the ZPL at the expense of collection count rates and hence measurement duration. The collection enhancement offered by a CBG structure is therefore desirable to offset the reduction in brightness due to spectral filtering.

The kinetic spectrum plot in Figure 4b shows that the spectral stability is very high, with no ZPL wavelength deviation over two minutes observed within the limit of the spectrometer. Minimizing spectral diffusion is also crucial for advanced quantum optics experiments involving photon interference [38-40], and this result reflects the minimal impact that the fabrication procedure has on the quality of the hBN lattice surrounding the B-center. Additional spectral stability may be due to the reduced crystal volume of the central CBG disk, leading to a lower population of neighboring charge traps whose fluctuations cause inhomogeneous linewidth broadening [41]. We note that the ZPL occurs at 432.5 nm, which is moderately far from the expected value of 436.0 nm ± 0.2 nm [42], likely due to anomalous local strain conditions. From the power saturation measurement in Figure 4c we estimate a maximum photon count rate of 18 kcps, and the saturation power occurs at 0.4 mW. The second order correlation measurement in Figure 4d indicates that the emission is predominantly from a single emitter, with the small degree of bunching evident in the shoulders being due to excitation above the saturation power.

In summary, we explored etching options for monolithic hBN fabrication in order to produce monolithic circular Bragg grating cavities. We positioned single quantum emitters within the fabricated cavities and demonstrated a 6-fold emission collection enhancement of the 436 nm emission compared to uncoupled emitters in bulk hBN. We discuss the effects of design scaling and EBL conditions on the fabrication and photonic function of the devices and observe spectrometer-limited temporal spectral stability for a coupled emitter at 5 K.

**Acknowledgments**
We thank Angus Gale for assistance with the electron irradiation. The authors acknowledge financial support from the Australian Research Council (CE200100010, FT220100053) and the Office of Naval Research Global (N62909-22-1-2028). The UTS node of the ANFF is greatly acknowledged for access to nanofabrication tools.

**Methods**
**Flake Preparation**

Flakes of pristine and carbon doped hBN were mechanically exfoliated with scotch tape onto 285 nm thick $SiO_2$ on Si. Flakes were identified as appropriate for fabrication using optical contrast and were 50 μm by 50 μm in area and approximately 275 nm thick.

**Patterning**

Samples were prepared for EBL with spin coating of a positive resist. Either CSAR, for 10 s at 900 rpm and 50 s at 4500 rpm, or PMMA, for 10 s at 800 rpm and 50 s at 3000 rpm. They were then baked for three minutes at 180 °C. Patterning was done with a RAITH EBL system in an FEG-SEM (Zeiss Supra 55 VP). The CSAR samples were patterned with an electron beam energy of 30 kV, beam current of 20 pA and fluence of 110 $\mu C/cm^2$ was used. The PMMA samples were patterned with an electron beam energy of 30 kV, beam current of 40 pA and fluence of 200 $\mu C/cm^2$. For a robust fabrication, a pattern was designed for an array of devices scaled by 2% so the Bragg coefficients for each column ranged from 240.64 nm in the first (left) column to 256 nm in the fourth and 266.24 nm in the 6th (right). In addition, each row was exposed with a different electron fluence during EBL. This centered around 110 $\mu C/cm^2$ and ranged from 104.5 $\mu C/cm^2$ in the bottom row to 115.5 $\mu C/cm^2$ in the top for the CSAR samples.

**Etching**

After development of the EBL the patterns were transferred into the hBN using ion beam etching, chemically assisted ion beam etching and reactive ion etching. IBE and CAIBE were conducted in an Intlvac Nanoquest I with a PMMA mask. IBE was conducted for 5 min, with an argon beam energy of 400 V. CAIBE was conducted for 10 min with an argon beam energy of 300 V and a flow of 5 sccm of $SF_6$ at the sample surface. Chlorine- and fluoride-based RIE were conducted in separate TRION ICP Plasma Chambers. Both RIE samples used CSAR EBL resist as a hard mask and were oxygen plasma cleaned for 5 s before etching. The chlorine sample was etched for 120 s at 5 mT with 5 W ICP and 100 W RIE power and 10 sccm of argon and 2 sccm of chlorine. The fluoride sample was etched for 37 s at 5 mT with 1 W ICP and 300 W RIE power, 60 sccm of argon and 1 sccm of $SF_6$. The CSAR resist was then chemically removed.

**Defect Creation**

Samples were spot irradiated in a FEI DB235 Dual Beam FIB/SEM with a 5 kV electron beam at 1.6 nA to create B-center emitters at multiple locations on the fabricated flake.

**Characterisation**

Samples were first optically characterized at room temperature on a lab-built scanning confocal microscope with a 405 nm continuous-wave (CW) laser (PiL040X, A.L.S. GmbH) and an XYZ piezostage (NanoCube P-611.3). The sample was excited and emission was collected through a 0.9 NA Nikon objective. Collection was then filtered through a long pass 405 nm dichroic mirror and a 460 nm band pass filter. It was coupled into a 50:50 fiber recorded by avalanche photodiode single-photon detector (Excelitas Technologies) and spectrometer (Princeton Instruments, Inc.). Cryogenic characterization was performed on a closed-loop cryostat (Attocube) using a 0.82 NA objective (Attocube LT-APO/VIS/0.82).

**Photonic Simulation**

3D finite-domain time-difference (FDTD) method is used. The cavity is made of a hBN layer with a thickness of 200 nm that is on SiO$_2$ substrate. In this setup, starting from the central hBN disk with a diameter of 512 nm, each successive ring has a width and spacing of 256 nm and 97 nm, respectively. The simulation utilises a refractive index of 2.2 for in-plane and 1.9 for out-of-plane.

**References**

[1] L. S. Madsen et al., *Quantum Computational Advantage with a Programmable Photonic Processor*, Nature **606**, 7912 (2022).
[2] Z.-D. Li et al., *Experimental Quantum Repeater without Quantum Memory*, Nat. Photonics **13**, 644 (2019).
[3] Z. A. Kudyshev, D. Sychev, Z. Martin, O. Yesilyurt, S. I. Bogdanov, X. Xu, P.-G. Chen, A. V. Kildishev, A. Boltasseva, and V. M. Shalaev, *Machine Learning Assisted Quantum Super-Resolution Microscopy*, Nat. Commun. **14**, 1 (2023).
[4] S. Slussarenko and G. J. Pryde, *Photonic Quantum Information Processing: A Concise Review*, Appl. Phys. Rev. **6**, 041303 (2019).
[5] C. Couteau, S. Barz, T. Durt, T. Gerrits, J. Huwer, R. Prevedel, J. Rarity, A. Shields, and G. Weihs, *Applications of Single Photons to Quantum Communication and Computing*, Nat. Rev. Phys. **5**, 326 (2023).
[6] C. Becher et al., *2022 Roadmap for Materials for Quantum Technologies*, 38 (n.d.).
[7] I. Aharonovich, J.-P. Tetienne, and M. Toth, *Quantum Emitters in Hexagonal Boron Nitride*, Nano Lett. **22**, 9227 (2022).
[8] M. Turunen, M. Brotons-Gisbert, Y. Dai, Y. Wang, E. Scerri, C. Bonato, K. D. Jöns, Z. Sun, and B. D. Gerardot, *Quantum Photonics with Layered 2D Materials*, Nat. Rev. Phys. **4**, 219 (2022).
[9] N. Ronceray et al., *Liquid-Activated Quantum Emission from Pristine Hexagonal Boron Nitride for Nanofluidic Sensing*, Nat. Mater. 1 (2023).
[10] A. R.-P. Montblanch, M. Barbone, I. Aharonovich, M. Atatüre, and A. C. Ferrari, *Layered Materials as a Platform for Quantum Technologies*, Nat. Nanotechnol. **18**, 6 (2023).
[11] I. Zhigulin, J. Horder, V. Ivady, S. J. U. White, A. Gale, C. Li, C. J. Lobo, M. Toth, I. Aharonovich, and M. Kianinia, *Stark Effect of Blue Quantum Emitters in Hexagonal Boron Nitride*, Phys. Rev. Appl. **19**, 044011 (2023).
[12] G. Grosso, H. Moon, B. Lienhard, S. Ali, D. K. Efetov, M. M. Furchi, P. Jarillo-Herrero, M. J. Ford, I. Aharonovich, and D. Englund, *Tunable and High-Purity Room Temperature Single-Photon Emission from Atomic Defects in Hexagonal Boron Nitride*, Nat. Commun. **8**, 705 (2017).
[13] T. Vogl, M. W. Doherty, B. C. Buchler, Y. Lu, and P. K. Lam, *Atomic Localization of Quantum Emitters in Multilayer Hexagonal Boron Nitride*, Nanoscale **11**, 14362 (2019).
[14] B. Shevitski et al., *Blue-Light-Emitting Color Centers in High-Quality Hexagonal Boron Nitride*, Phys. Rev. B **100**, 155419 (2019).
[15] C. Fournier et al., *Position-Controlled Quantum Emitters with Reproducible Emission Wavelength in Hexagonal Boron Nitride*, Nat. Commun. **12**, 3779 (2021).
[16] A. Gale, C. Li, Y. Chen, K. Watanabe, T. Taniguchi, I. Aharonovich, and M. Toth, *Site-Specific Fabrication of Blue Quantum Emitters in Hexagonal Boron Nitride*, ACS Photonics **9**, 2170 (2022).
[17] N. Chejanovsky et al., *Structural Attributes and Photodynamics of Visible Spectrum Quantum Emitters in Hexagonal Boron Nitride*, Nano Lett. **16**, 7037 (2016).
[18] E. Glushkov et al., *Engineering Optically Active Defects in Hexagonal Boron Nitride Using Focused Ion Beam and Water*, ACS Nano **16**, 3695 (2022).
[19] R. N. Patel, D. A. Hopper, J. A. Gusdorff, M. E. Turiansky, T.-Y. Huang, R. E. K. Fishman, B. Porat, C. G. Van de Walle, and L. C. Bassett, *Probing the Optical Dynamics of Quantum Emitters*

*in Hexagonal Boron Nitride*, PRX Quantum **3**, 030331 (2022).

[20] S. Castelletto, F. A. Inam, S. Sato, and A. Boretti, *Hexagonal Boron Nitride: A Review of the Emerging Material Platform for Single-Photon Sources and the Spin–Photon Interface*, Beilstein J. Nanotechnol. **11**, 740 (2020).

[21] S. Ates, L. Sapienza, M. Davanco, A. Badolato, and K. Srinivasan, *Bright Single-Photon Emission From a Quantum Dot in a Circular Bragg Grating Microcavity*, IEEE J. Sel. Top. Quantum Electron. **18**, 1711 (2012).

[22] L. Sapienza, M. Davanço, A. Badolato, and K. Srinivasan, *Nanoscale Optical Positioning of Single Quantum Dots for Bright and Pure Single-Photon Emission*, Nat. Commun. **6**, 1 (2015).

[23] H. Wang et al., *On-Demand Semiconductor Source of Entangled Photons Which Simultaneously Has High Fidelity, Efficiency, and Indistinguishability*, Phys. Rev. Lett. **122**, 113602 (2019).

[24] M. Moczała-Dusanowska, Ł. Dusanowski, O. Iff, T. Huber, S. Kuhn, T. Czyszanowski, C. Schneider, and S. Höfling, *Strain-Tunable Single-Photon Source Based on a Circular Bragg Grating Cavity with Embedded Quantum Dots*, ACS Photonics **7**, 3474 (2020).

[25] S. Xia, T. Aoki, K. Gao, M. Arita, Y. Arakawa, and M. J. Holmes, *Enhanced Single-Photon Emission from GaN Quantum Dots in Bullseye Structures*, ACS Photonics **8**, 1656 (2021).

[26] N. Livneh, M. G. Harats, D. Istrati, H. S. Eisenberg, and R. Rapaport, *Highly Directional Room-Temperature Single Photon Device*, Nano Lett. **16**, 2527 (2016).

[27] H. Abudayyeh et al., *Single Photon Sources with near Unity Collection Efficiencies by Deterministic Placement of Quantum Dots in Nanoantennas*, APL Photonics **6**, 036109 (2021).

[28] L. Li, E. H. Chen, J. Zheng, S. L. Mouradian, F. Dolde, T. Schröder, S. Karaveli, M. L. Markham, D. J. Twitchen, and D. Englund, *Efficient Photon Collection from a Nitrogen Vacancy Center in a Circular Bullseye Grating*, Nano Lett. **15**, 1493 (2015).

[29] S. Kumar, C. Wu, D. Komisar, Y. Kan, L. F. Kulikova, V. A. Davydov, V. N. Agafonov, and S. I. Bozhevolnyi, *Fluorescence Enhancement of a Single Germanium Vacancy Center in a Nanodiamond by a Plasmonic Bragg Cavity*, J. Chem. Phys. **154**, 044303 (2021).

[30] N. M. H. Duong et al., *Enhanced Emission from WSe$_2$ Monolayers Coupled to Circular Bragg Gratings*, ACS Photonics **5**, 3950 (2018).

[31] Y. Liu et al., *Photoluminescence Enhancement of InSe by Coupling with Circular Bragg Grating*, Laser Photonics Rev. 2300234 (2023).

[32] J. E. Fröch, L. P. Spencer, M. Kianinia, D. D. Totonjian, M. Nguyen, A. Gottscholl, V. Dyakonov, M. Toth, S. Kim, and I. Aharonovich, *Coupling Spin Defects in Hexagonal Boron Nitride to Monolithic Bullseye Cavities*, Nano Lett. **21**, 6549 (2021).

[33] D. Gérard, M. Rosticher, K. Watanabe, T. Taniguchi, J. Barjon, S. Buil, J.-P. Hermier, and A. Delteil, *Top-down Integration of an hBN Quantum Emitter in a Monolithic Photonic Waveguide*, Appl. Phys. Lett. **122**, 264001 (2023).

[34] M. Nonahal, J. Horder, A. Gale, L. Ding, C. Li, M. Hennessey, S. T. Ha, M. Toth, and I. Aharonovich, *Deterministic Fabrication of a Coupled Cavity–Emitter System in Hexagonal Boron Nitride*, Nano Lett. **23**, 6645 (2023).

[35] S. Kim, J. E. Fröch, J. Christian, M. Straw, J. Bishop, D. Totonjian, K. Watanabe, T. Taniguchi, M. Toth, and I. Aharonovich, *Photonic Crystal Cavities from Hexagonal Boron Nitride*, Nat. Commun. **9**, 1 (2018).

[36] M. Nonahal, C. Li, H. Ren, L. Spencer, M. Kianinia, M. Toth, and I. Aharonovich, *Engineering Quantum Nanophotonic Components from Hexagonal Boron Nitride*, Laser Photonics Rev. **17**, 2300019 (2023).

[37] J. Iles-Smith, D. P. S. McCutcheon, A. Nazir, and J. Mørk, *Phonon Scattering Inhibits Simultaneous Near-Unity Efficiency and Indistinguishability in Semiconductor Single-Photon Sources*, Nat. Photonics **11**, 521 (2017).

[38] C. Fournier, S. Roux, K. Watanabe, T. Taniguchi, S. Buil, J. Barjon, J.-P. Hermier, and A. Delteil, *Two-Photon Interference from a Quantum Emitter in Hexagonal Boron Nitride*, Phys. Rev. Appl. **19**, L041003 (2023).

[39] A. Sipahigil, K. D. Jahnke, L. J. Rogers, T. Teraji, J. Isoya, A. S. Zibrov, F. Jelezko, and M. D. Lukin, *Indistinguishable Photons from Separated Silicon-Vacancy Centers in Diamond*, Phys. Rev. Lett. **113**, 113602 (2014).

[40] B. Kambs and C. Becher, *Limitations on the Indistinguishability of Photons from Remote Solid*

*State Sources*, New J. Phys. **20**, 115003 (2018).
[41] J. Wolters, N. Sadzak, A. W. Schell, T. Schröder, and O. Benson, *Measurement of the Ultrafast Spectral Diffusion of the Optical Transition of Nitrogen Vacancy Centers in Nano-Size Diamond Using Correlation Interferometry*, Phys. Rev. Lett. **110**, 027401 (2013).
[42] J. Horder, S. J. U. White, A. Gale, C. Li, K. Watanabe, T. Taniguchi, M. Kianinia, I. Aharonovich, and M. Toth, *Coherence Properties of Electron-Beam-Activated Emitters in Hexagonal Boron Nitride Under Resonant Excitation*, Phys. Rev. Appl. **18**, 064021 (2022).